# Simulation of GaN-Based Light Emitting Diodes Incorporating Composition Fluctuation Effects

Sheikh Ifatur Rahman[1*], Zane Jamal-Eddine[1], Zhanbo Xia[1], Mohammad Awwad[1], Rob Armitage[2], and Siddharth Rajan[1,3*]

[1]Department of Electrical and Computer Engineering, The Ohio State University, Columbus, Ohio 43210, USA.
[2]Lumileds LLC, San Jose, California, 95131, USA.
[3]Department of Materials Science and Engineering, The Ohio State University, Columbus, Ohio 43210, USA.

* Corresponding author(s)
Corresponding author(s) email: rahman.227@osu.edu; rajan.21@osu.edu

**Abstract:**

III-Nitride light emitting diodes (LEDs) are widely used in a range of high efficiency lighting and display applications, which have enabled significant energy savings in the last decade. Despite the wide application of GaN LEDs, transport mechanisms across InGaN/GaN heterostructures in these devices are not well-explained. Fixed polarization sheet charges at InGaN/GaN interfaces lead to large interface dipole charges, which create large potential barriers to overcome. One-dimensional models for transport across such heterostructures predict turn-on voltages that are significantly higher than that found in real devices. As a result, conventional models for transport cannot predict the performance of new designs such as for longer wavelength LEDs, or for multi-quantum well LEDs. In this work, we show that incorporating low and high Indium compositions within quantum wells at the submicron scale can provide accurate prediction of the characteristics of GaN/InGaN light emitting diodes.

**Introduction:**

III-Nitride light emitting diodes and diode lasers are a key technology for a wide range of lighting, display, and communication applications due to their efficiency and manufacturability.[1] Currently, research efforts are focused on realizing higher power density, longer wavelength emission, multi-active region emission, and greater functionality. Despite the considerable commercial success of LEDs across various wavelength regimes, the electrical characteristics and turn-on voltage are still relatively poorly understood and are not well-explained by standard device models. Simulation of LEDs using standard device models greatly over-estimates the turn-on voltage of LEDs. As LEDs are integrated into various applications such as micro-LED displays and light-based communications, it becomes increasingly important to have robust predictive models that explain the electrical characteristics. Furthermore, such predictive models may be used for future designs of LEDs too.

Early insight into the reason for the anomalously low turn-on voltage was given by Wu and Speck, who showed through detailed materials characterization and modeling, that compositional fluctuations in the (In,Ga)N alloy in the quantum well can cause significant changes in carrier transport and emission properties.[2-6] Using a "quantum landscape" formalism [5, 6], they were able to qualitatively explain the lower turn-on voltage in III-Nitride (In,Ga)N/GaN-based MQW LEDs, though exact quantitative agreement with experimental data was still not obtained. Key transport mechanisms such as carrier tunneling paths or any tunneling phenomenon are not defined in those models which plays vital role for interwell carrier transport. Atom probe measurements on (In,Ga)N thin films and (In,Ga)N QWs have showed presence of In segregation causing non-uniform indium fluctuation in both lateral and vertical direction which significantly impact device electrical and optical properties.[3, 7-20] Galtrey et. al. showed through a 3D

APT study that Indium in (In,Ga)N for a MQW structure is randomly placed with the composition frequency distribution ranging from 5% (In,Ga)N to 33% (In,Ga)N for an average 19% (In,Ga)N QW structure [21]. Mehrtens et. al. through both HAADF-STEM and APT analysis showed that for (In,Ga)N QWs with composition greater that 25% the highest composition obtained ranges to 31% with lowest around 14% [15]. Impact of such alloy compositional fluctuations is significant in light emitting devices as the device optical behavior as well the electrical performance is heavily dependent on these local variations of the compositions. The carriers, both electron and holes when injected into the active region QW get localized in the high In composition region due to deep localized band-edge potential from the compositional fluctuations.[22] The quantum confinement in the in-plane potential wells will generate localized energy states which will vary due to Indium fluctuation changing the lateral sizes of in-plane wells broadening the emission spectrum. [22] The polarization field in the QW also gets screened by increasing local current density at high carrier injection condition. Many of these effects, which impact the device at the nanometer scale, are ignored in standard LED device simulations when using uniform-composition quantum wells. Recently, a comparative analysis between 1D, and 3D simulation based on landscape theory incorporated the effect of alloy fluctuations in the active region and explained the source of extra voltage drop for green (In,Ga)N LEDs and the sequential injection of carriers in the MQW structure.[6] Similar study on both blue and green (In,Ga)N LED showed the effect of polarization induced barrier and large band offsets for carrier injection into the active region.[5] While the 3D simulation computes the turn on voltage and JV characteristics better than the 1D model, both the model over-estimates the turn on voltage compared to the experimental value. While the reports succeeded in explaining the extra turn on voltage for green LED through the inclusion of random alloy fluctuation in the model, fewer details were reported in terms of injection site for carriers, tunneling of carriers through multi-quantum well structures, and lateral carrier transport. 3D simulation model using only random alloy fluctuation predicted forward voltage closer to experimental value but showed increasing turn on and forward voltage drop for increasing number of QWs which is usually not seen in experimental devices. Later, simulation with a combination of random alloy fluctuation and V-pits injection mechanism (works as alternative path for carrier injection though semi-polar and non-polar planes along with vertical injection) showed lower forward voltage drop estimation for green LEDs [23].

The key parts of a simplified typical (In,Ga)N/GaN LED structure are shown in figure 1(a). The LED consists of an n-type GaN layer, GaN/(In,Ga)N multi-quantum (in this case 3) well regions where (In,Ga)N is the active light-emitting layer, an electron blocking layer, and a top p-GaN layer. A simulation of the energy band diagram (using Silvaco) of this 1-dimensional structure at equilibrium is shown in figure 1(b). The experimental and simulated electrical characteristics of this structure are shown in figure 1(c). The simulated IV curve shows higher forward voltage drop due to polarization discontinuities at the (In,Ga)N/GaN heterojunctions, which lead to large electric fields within the MQW region. This leads to the formation of increased potential barriers between quantum wells and at the edge of the MQW region (shown in figure 1 (b)) with the top p-type and bottom n-type injection layers. To turn on the LED and populate the quantum wells with carriers, sufficient bias must be applied to overcome these depletion barriers. Thus, the theoretically expected turn-on voltage for such a device is significantly larger than the band gap (shown in figure 1 (c)) and higher than experiment (which is very close to the bandgap of the quantum well layers). The experimental results suggest that the potential barriers predicted by the simulated energy band diagram are not in fact preventing carrier injection into the quantum well.

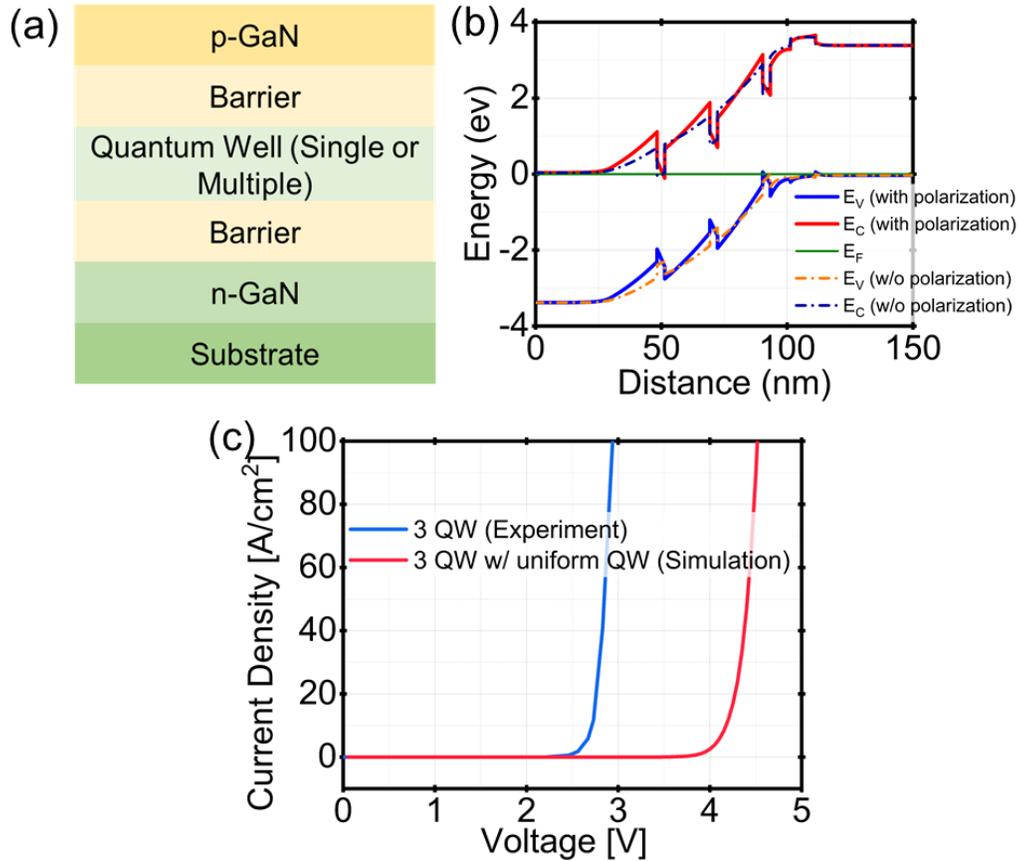

**Figure 1**. (a) Convention LED epitaxial structure, (b) Equilibrium band diagram of a 3 QW LED with 25% Indium content only, (c) JV characteristics of a 3QW green LED from experiment and simulated uniform QW Indium content.

In this work, we report on incorporation of Indium compositional fluctuations using an industry-standard device simulation tool Silvaco, which incorporates semi-classical transport mechanisms such as thermionic and field emission. The model is found to have very good agreement with the experimental data. Furthermore, it gives insight into carrier transport physics at low and high carrier injections, and into vertical and in-plane carrier transport into the active region.

**Modeling:**

Fluctuation in the well region is imitated by two different (In,Ga)N composition placed side by side. Using a combination of two different (In,Ga)N compositions, this model shows sequential injection inside the multi quantum well active region, reveals the vertical injection mechanism for carriers through low In composition area as well as the lateral movement of carriers inside the quantum well, predicts electrical performance closer to the experimental data and follows forward voltage drop trends for MQW active regions.

The epitaxial structure of the devices investigated here are shown in figure 2(a). The information about the epitaxial structures was deduced from calibrations. The structures were grown by MOCVD on c-plane Si doped GaN templates. The device active region was varied (number of quantum wells = 2,3 and 4). The barriers on each side of the QWs were 18 nm thick with Si doping, except for the top-most barrier which was not doped and was capped with a 10 nm Mg-doped $Al_{0.15}Ga_{0.85}N$ electron blocking layer. The active region consisted of (In,Ga)N QWs (~3 nm). Devices similar to experimental structure were simulated in Silvaco TCAD software. The interface polarization at each heterojunction was calculated assuming bulk values [24-26] (Table 1). The quantum well regions were defined as optical emission regions, and optical/recombination constants were provided (radiative recombination coefficient $B_0=2\times10^{-11}$ cm$^3$/s [27-29], intrinsic auger recombination coefficient $C_0=2\times10^{-30}$ cm$^6$/s [29, 30]) (details in Table 2). A three-band wurtzite model based on k.p modeling accounts for the optical transitions between the conduction band and heavy/light/split-off hole valence bands. In this model, GaN/(In,Ga)N heterojunctions are treated as abrupt interface and thermionic emission transport model (TE), field emission transport model (FE) and non-local quantum barrier tunneling model are placed to calculate the total current through the device. Field emission model calculates the tunneling current based on the value of electric field at the interface. The non-local quantum barrier tunnel model (QB tunnel) is placed across the quantum wells and quantum barriers which evaluates the tunneling current at all energies at which tunneling is possible and converts into a recombination or generation rate and inserts into current continuity equation. The model does not include any other alternate transport mechanism like V-pits injection.

Table 1. Piezoelectric and spontaneous polarization parameters for GaN, AlN and InN [24-26]:

|  | $e_{31}$ (C/m$^2$) | $e_{33}$ (C/m$^2$) | a (Å) | $C_{13}$ (GPa) | $C_{33}$ (GPa) | Psp (C/m$^2$) |
|---|---|---|---|---|---|---|
| GaN | -0.49 | 0.73 | 3.189 | 103 | 405 | -0.029 |
| AlN | -0.6 | 1.46 | 3.112 | 108 | 373 | -0.081 |
| InN | -0.57 | 0.97 | 3.54 | 92 | 224 | -0.032 |

Table 2. Different key material properties in the model:

| Layer | Mobility (electron/hole) (cm$^2$/Vs) | Relative electron/hole effective mass | SRH lifetime (electron/hole) (second) | Optical recombination rate (cm$^3$/s) | Auger coefficients (cm$^6$/s) |
|---|---|---|---|---|---|
| GaN | 300/10 | 0.19 m$_e$ / 0.8 m$_h$ | $1\times10^{-9}$ | $1.1\times10^{-8}$ | $1\times10^{-34}$ |
| (In,Ga)N |  | 0.19 m$_e$ / 0.8 m$_h$ | $5\times10^{-7}$ | $2\times10^{-11}$ | $2\times10^{-30}$ |
| AlGaN | 250/5 | - | $1\times10^{-9}$ | $1.1\times10^{-8}$ | $1\times10^{-34}$ |

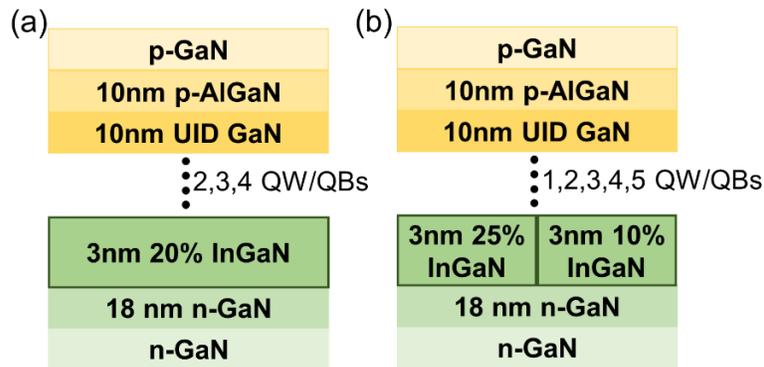

**Figure 2.** (a) Epitaxial Structure of Experimental Device (b) Simulated Green LED device structure.

To incorporate alloy fluctuations, a simplified model as shown in figure 2(b) was used. A high composition region (corresponding to the emission wavelength) and low composition region inside the quantum well were included as adjacent regions within the quantum well. To match the experimental LEDs (emitting in the green wavelength range), the In mole-fraction for the "high" In-content region was assumed to be 25%, and that for the low In-content region was assumed to be 10% placed side by side with equal dimension. The high and low composition In-content values were based on atom probe tomography reports and compositional mapping reports – the composition values for the "high" and "low" regions were chosen by selecting the highest and lowest compositions from these experimental reports.[4, 17, 21, 31-33] The simulation was run on a structure with in-plane dimension of 100 nm.

**Results and Discussion:**

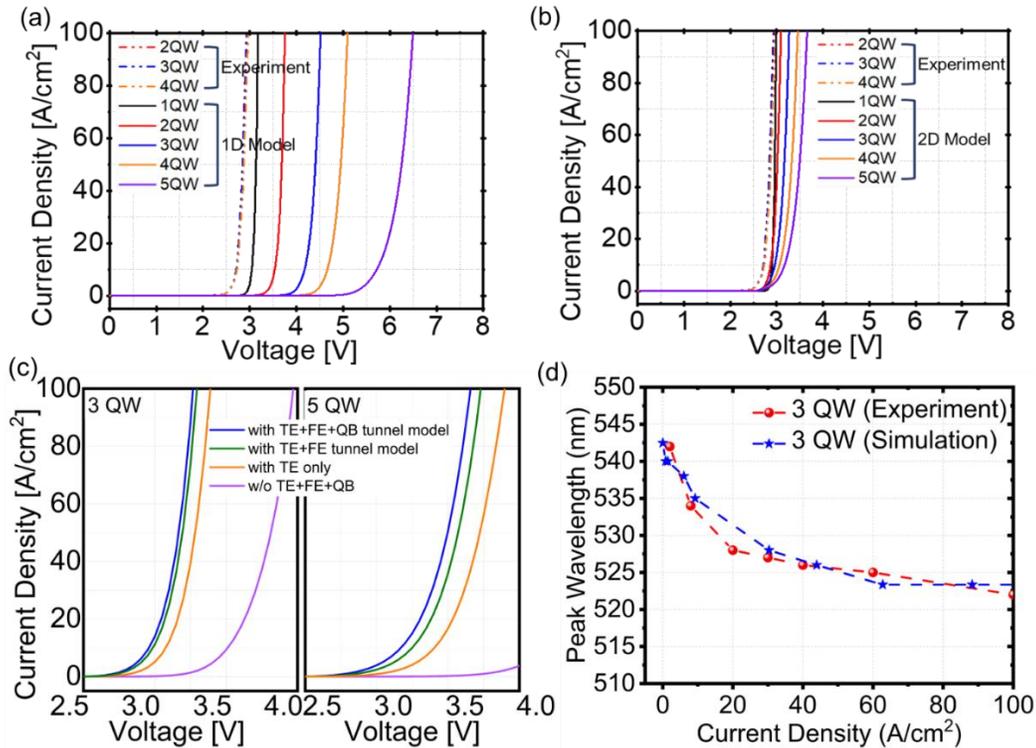

**Figure 3.** a) Simulated characteristics of green LED devices with uniform composition quantum well (solid lines). b) Simulated characteristics of green LED devices with a two-composition structures (solid lines). (c) Simulated characteristics of green LED devices with and without different transport models for MQW structures. (d) Wavelength shift obtained from the electroluminescence measurement of experimental 3QW device and simulated 3QW device.

The electrical characteristics from experiment and simulation for the uniform well are shown in figure 3(a), and these show significant overestimation of the turn-on voltage and voltage drop in all cases despite the inclusion of all the relevant transport models. Figure 3(b) shows electrical characteristics from the non-uniform QW model. Clearly the inclusion of the low-composition regime as a parallel current path shows significantly better agreement with the experimental data, both in terms of the turn-on voltage and in terms of the shape of the current-voltage characteristics. For example, the experimental data shows that the turn-on voltage is almost independent of the number of quantum wells. The uniform model shows a large increase in the turn-on voltage with quantum well number, but the in the case of the non-uniform model, the voltage drop at low current densities (< 5 A/cm$^2$) agrees quite well with experiment due to the presence of low In-content region which has lower polarization charges and thus lower electrostatic barrier. Figure 3(c) displays the simulated forward electrical characteristics of MQW LED devices showing the improvement of forward voltage prediction through the inclusion of different transport mechanism. When simulated without thermionic emission and any tunneling models, MQW structures shows extra voltage drop of 0.6 V for 3 QW and 1.1 V for 5 QW structures at 20 A/cm$^2$. The inclusion of the thermionic model reduces the forward voltage drop significantly for MQW structures but still underestimates due to absence of the possible tunneling transport mechanisms. Inclusion of the FE tunnel and QB tunnel model account for the possible tunneling mechanism at the depletion barrier and through the quantum barriers during interwell carrier transport respectively. The importance of including the tunnel models becomes more significant with increasing number of QWs as without such models the estimated forward voltage deviates from the trend of the experimental devices. Figure 3(d) shows experimental and simulated wavelength shift obtained from a 3QW structure using the model used here. The simulated emission at low current peaks at ~ 543 nm, and blue-shifts to 523 nm at 100 A/cm$^2$. The wavelength shift of the experimental 3QW device matches well with the simulated data.

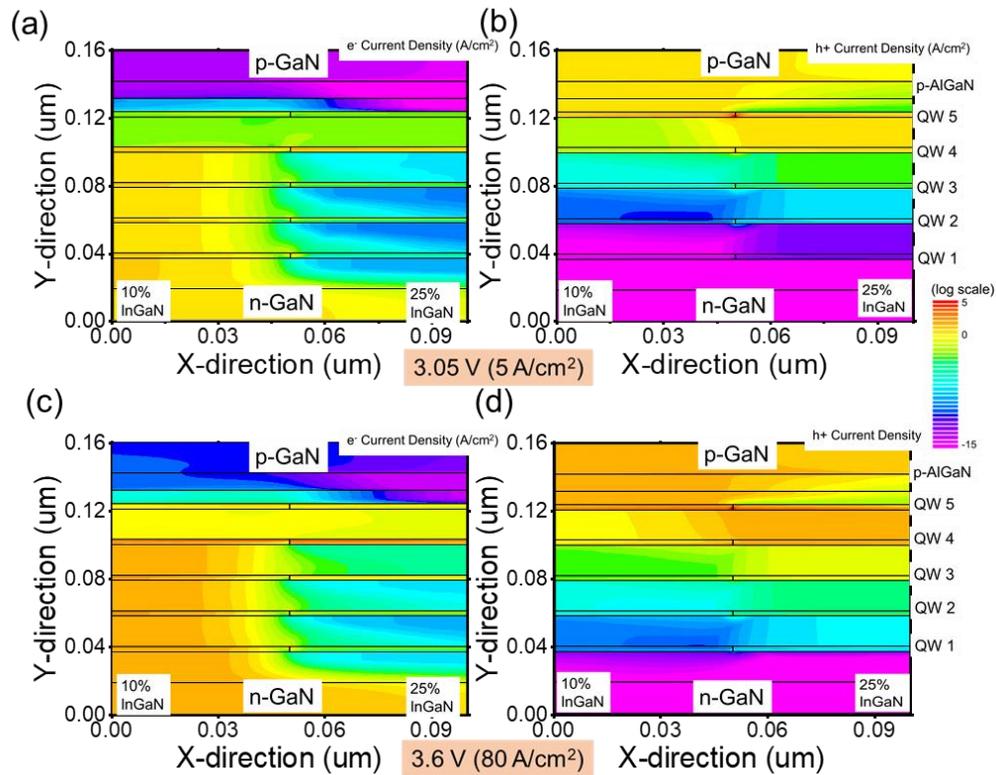

**Figure 4.** Contour maps at the active region showing electron current density ((a), (c)) and hole current density ((b), (d)) when the device is operated at 5 A/cm$^2$ and 80 A/cm$^2$ respectively.

The critical role of current injection and lateral carrier transport are discussed below to further explain these characteristics. The low In-content region in the non-uniform LEDs have lower polarization fields and lower band offsets, leading to lower potential barrier for electron and hole injection. Figure 4 shows the electron and hole current density at low current injection and high current injection regime for a 5QW active region device. It can be seen from this that the electrons and holes are injected through the low composition (In,Ga)N region. Electron current density contour plots (figure 4(a), (c)) show that electrons get injected through the low In-content region while the high In-content block the electron injection vertically due to higher electrostatic barrier. The potential barrier caused by polarization field in low In-content region is smaller compared to the high In-content region allowing the electron to flow through the low In region and localize at high In region. The low In-content region with lower band offsets allocates early overshoot of carriers allowing the movement of carriers to high In region laterally. Since all the carriers (both electron and hole are localized) gets localized in the high In-content region, majority of the recombination will take place in the high In region allowing the emission wavelength to be corresponding to the transition energy of the high In region. Electrons with lower effective mass can transport through the low In-content faster than the heavier holes, reaching the QW closer to the p-region whereas holes require higher bias to overcome the polarization induced barrier at each QW. At higher current densities and increasing forward voltages MQW structures in the simulated devices shows higher resistance as the interwell carrier transport is governed by thermionic emission of holes [34]. Uniform carrier distribution is not observed due to low carrier densities across all the QWs showing that these (In, Ga)N/GaN LEDs are limited by the hole transport for efficient performance.

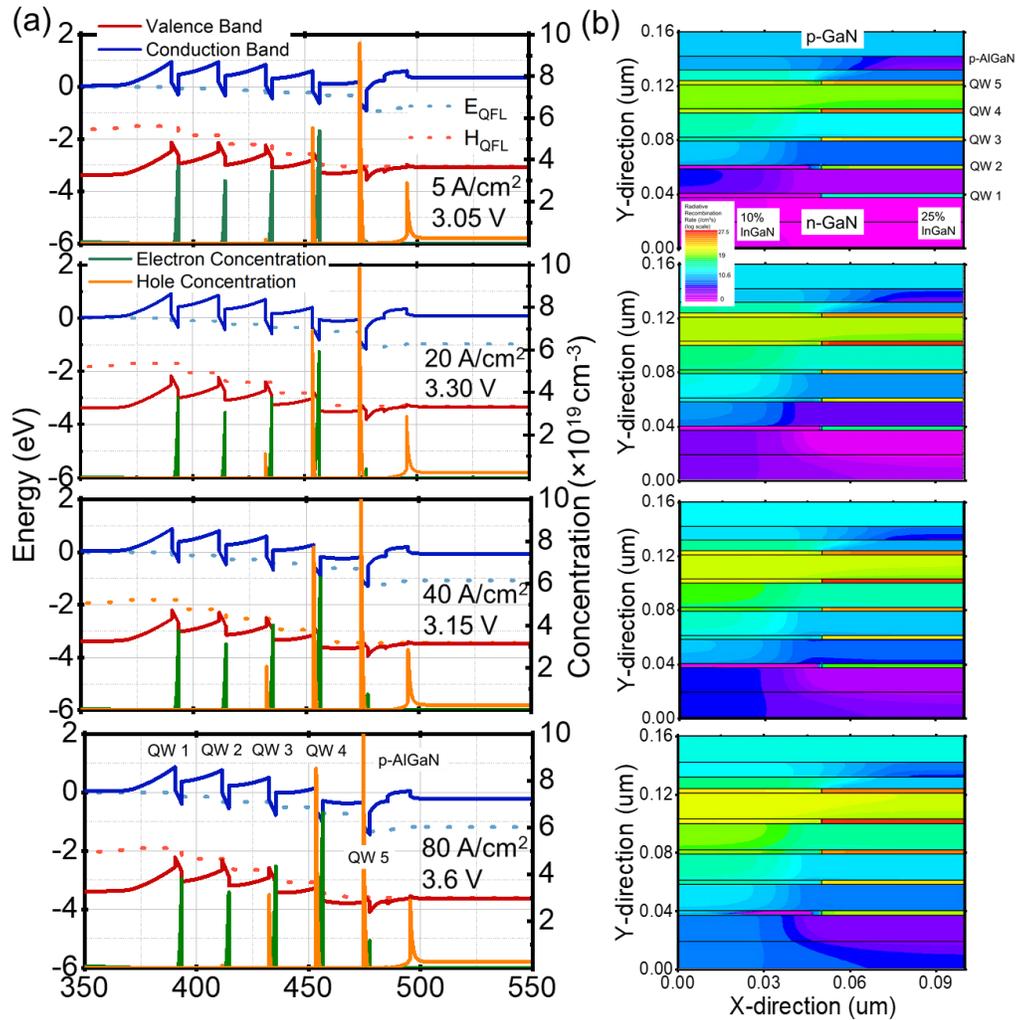

**Figure 5.** (a) Sequential injection of electrons and holes in MQW structure with increasing current injection at 5 A/cm$^2$, 20 A/cm$^2$, 40 A/cm$^2$ and 80 A/cm$^2$. Band diagrams and electron and hole concentrations shown are only for the high In-content region only. (b) Corresponding contour maps of the radiative recombination

sites at the same current density in figure (a) shows majority radiative recombination at the high In-content region.

Figure 5 shows the distribution of electrons and holes at different current densities in the high In-content region for a 5-QW active region structure. As expected, the hole density is high in the wells near the p-type and drops rapidly in the bottom three wells (adjacent to the n-type layer). Also, while the electron density is more uniformly distributed, it is still lower than the hole density in the wells near the p-GaN layer. The simulations also show that the transport of electrons from the n-side to the wells is mainly through the low In-content region. The transport of holes occurs through both the high and low In-content region in the QW closer to the p-region. The radiative recombination contour (Figure 5(b)) shows most of the emission occurring in the high In-content region as both electrons and holes are confined in those regions. The radiative recombination is limited to the QW close to the p-GaN side of the device. The contour map of electron current density (figure 4(c)) at 80 A/cm$^2$ show that electrons flow through the low In-content regions before recombining at the QWs close to the p-side. Negligible recombination is seen in the low In-content regions.

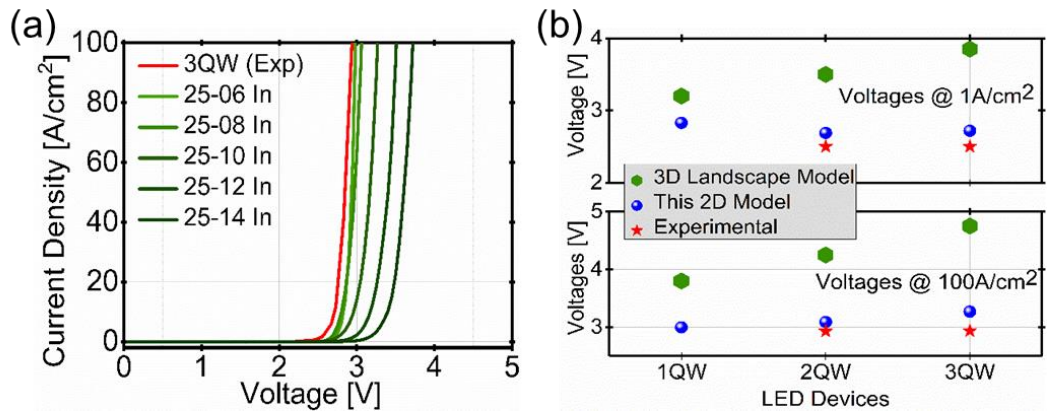

**Figure 6**. a) Simulated LED JV characteristics with varying low In-content of QWs in a 3QW device, b) shows the comparison of the improvement of prediction of forward voltage prediction at 1 A/cm$^2$ and 100 A/cm$^2$ when compared with the existing 3D model from [6].

In the models shown here, the low In-content region was chosen to be 10%, while the high In-content region was chosen to be 25% (to match emission characteristics). To further understand the impact of the assumed low In-content composition on LED characteristics, 3QW structure devices were simulated with varying low In content keeping the high In-content fixed at 25%. As the low In-content in the model is reduced the JV characteristics shows better fit with the experimental results (shown in figure 6 (a)). This suggests that the charge injection in real LEDs may be occurring in regions where the effective In-content is relatively low. Figure 6(b) shows a comparison of the predicted forward voltage drop at 1 A/cm$^2$ and 100 A/cm$^2$ for experimental devices, this non uniform QW model and reported 3D model using landscape theory [6]. The model presented here gives a good match with the experiment. We believe that the inclusion of compositional fluctuation and the relevant transport mechanisms (tunneling, thermionic emission) help to make the model used here accurate, even though the structure used (two-composition) is significantly simplified when compared with the actual random distribution of compositions.

**Conclusion:**

In conclusion, a simplified structure incorporating two Indium compositions in the quantum well region of an LED was used to predict the characteristics of III-Nitride LEDs. It was shown that 2-dimensional device simulations that use such a 2-composition model can accurately predict the IV and optical properties of real III-Nitride LEDs. The model provides insight into carrier transport within the LED, including the injection of carriers through low-composition regions, lateral transport of carriers within wells, and the recombination of carriers in the high-composition regions. The model presented here could provide opportunities for better analysis and design of III-Nitride LEDs in the future and enable extraction of circuit and behavioral models that are relevant for high-speed (Li-Fi) or micro-LED applications.

**Acknowledgements:**

This material is based upon the work supported by the U.S. Department of Energy's Office of Energy Efficiency and Renewable Energy (EERE) under the Building Technologies Office award no. DE-EE0009163. The views expressed in the article do not necessarily represent the views of the U.S. Department or the U.S. Government.